\documentclass[aps,pre,floats]{revtex4}
\usepackage{amsmath}
\usepackage{graphicx}

\begin{document}
\renewcommand{\topfraction}{.75}
\renewcommand{\bottomfraction}{.75}
\renewcommand{\textfraction}{0.2}
\renewcommand{\floatpagefraction}{0.7}

\title{Nonlinear lattice model of viscoelastic Mode III fracture}

\author{David A. Kessler}
\email{kessler@dave.ph.biu.ac.il}
\affiliation{Dept. of Physics, Bar-Ilan University, Ramat-Gan ISRAEL}
\author{Herbert Levine}
\email{levine@herbie.ucsd.edu}
\affiliation{Dept. of Physics, Univ. of Cal., San Diego, La Jolla, CA 92093-0319 USA}

\begin{abstract}
We study the effect of general nonlinear force laws in
viscoelastic lattice models of  fracture, focusing on the
existence and stability of steady-state Mode III cracks. We show
that the hysteretic behavior at small driving is very sensitive to the
smoothness of the force law.  At large driving, we find 
a Hopf bifurcation to a straight crack whose velocity is
periodic in time.  The frequency of the unstable bifurcating mode
depends on the smoothness of the potential, but is very close to
an exact period-doubling instability. Slightly above the onset of the
instability, the system settles into a exactly period-doubled
state, presumably connected to the aforementioned bifurcation
structure. We explicitly solve for this new state and map out its
velocity-driving relation.

\end{abstract}

\maketitle

\section{Introduction}
The problem of dynamic fracture has received increasing attention
in the physics community in the last decade~\cite{review}.  The
experiments of Fineberg, et. al.~\cite{fineberg}, showing
interesting dynamical behavior for large velocity cracks have been
at the center of this growing interest.  From a theoretical
point of view, the singularities at the crack tip associated with
continuum treatments of the crack problem make the problem
challenging.  The presence of these singularities necessitates a
treatment of the crack at the microscopic level, where the
continuum treatment and its associated singularities are not
applicable.  One line of attack on this problem, initiated by
Slepyan~\cite{slepyan}, has been through the study of lattice
models of cracks. These lattice models are simpler than full
atomistic simulations~\cite{mdsim} in that the connectivity of the
atoms is specified from the beginning, and so dislocations are
excluded.

Much progress has been made in understanding steady-state
propagation of cracks in lattice
systems~\cite{slepyan,mg,kl1,crack3,pk}. A key simplifying
assumption underlying much of this progress has been the
assumption of piecewise-linear forces between the particles, so
that the particles interact with Hookean springs, which break at
some critical extension, reducing the force to zero. With these
piecewise-linear interactions, the model admits an analytic
solution via the Wiener-Hopf technique.  This solution has been
carried out both for Mode I and Mode III cracks, for both finite
width and infinitely wide systems, with and without dissipation.

A general feature of these solutions is that they are problematic at both
very small and very large (roughly above 0.7 of the wave speed) velocities.
For small dissipation, the solutions are inconsistent at small velocities,
in that bonds on the crack surface which are assumed to crack at time $t=0$, say, in fact are seen to crack earlier. For large dissipation, the small
velocity solutions are consistent, but exhibit a backward (velocity increasing
with decreasing driving) velocity/driving curve, indicating the solutions
are unstable.  Thus, in any case, stable solutions are not found for small
velocities.

At large velocities, the analytic solutions are again inconsistent, this
time due to the breaking of additional bonds not on the crack surface.  The
analytic methods thus are unable to tell us anything about the dynamics beyond
this point.

An additional limitation of the piecewise-linear force law is that
it complicates the task of constructing a linear stability
treatment of the steady-state crack. This is due of course to the
discontinuous nature of the force law.  For these reasons, we
choose in this paper to examine steady-state crack propagation in
lattice models with arbitrary force-laws.  In particular, we study
a family of force-laws parameterized by $\alpha$, such that for
small $\alpha$ the force-law is smooth and the force-law goes over
to the piecewise-linear one in the limit of infinite $\alpha$. We
have previously studied~\cite{arrest} the behavior of arrested
Mode III cracks with this family of force-laws, and found that the
range of drivings for which arrested cracks exists narrows sharply
as $\alpha$ is reduced from infinity.  In this paper we examine
moving Mode III cracks for varying $\alpha$.  We find that the
behaviour for small velocity is very sensitive to $\alpha$.  For
large velocities, we find that the effect of finite $\alpha$ is to
convert the inconsistency to a regular linear instability, here of
Hopf type.  We show that at some distance into the unstable
regime, the system adopts a new form of steady-state behavior
which breaks the symmetry across the crack surface and has a
period-2 structure.  We conclude with some observations about the
model deep in the unstable regime.

\section{Model and General Methodology}
In this paper, we study a triangular lattice model of Mode III
cracking. We take the force exerted at site $\vec{x}_1$ by the
displacement at nearest neighbor site $\vec{x}_2$, with (scalar)
displacements $u_1$, $u_2$, respectively to be
\begin{equation}
f_{1,2}=(u_2-u_1)\frac{1 + \tanh(\alpha(1-\sigma_{1,2}(u_2-u_1)))}{1+\tanh(\alpha)}
\end{equation}
where $\alpha$ is a parameter which controls the smoothness of the
potential and $\sigma$ is positive if $\vec{x}_2$ lies to the left
of $\vec{x}_1$ or is further from the crack plane $y=0$.  This
form has the feature that the force goes to zero at large positive
extensions of the nearest-neighbor springs. In the limit of large
$\alpha$, this force-law approaches a step-function, (with
critical spring extension 1), a form introduced by
Slepyan\cite{slepyan}, and the subject of much
investigation\cite{slepyan2,marderliu,mg,kl1,crack3,pk}. The
lattice has $2N+2$ rows in the $y$-direction, separated by a
distance $\sqrt{3}/2$, so that the rows are labeled from
$y=-(2N+1)\sqrt{3}/4$ to $y=(2N+1)\sqrt{3}/4$. The displacements
$u$ of the bottom and top rows at $y=\pm(2N+1)\sqrt{3}/4$ are
constrained to be $\pm \Delta$.  We introduce a Kelvin-type
viscosity parameterized by $\eta$ via an additional dissipative
force
\begin{equation}
g_{1,2}=\eta  k_{1,2} \frac{d}{dt}(u_2 - u_1)
\end{equation}
where $k_{1,2}$ is an effective spring constant
\begin{equation}
k_{1,2}=f_{1,2}/(u_2 - u_1)
\end{equation}
This form was chosen so as to ensure a purely dissipative force
which in the limit of large $\alpha$ goes over to the form
$\eta\theta(1-(u_2-u_1))(\dot{u}_2-\dot{u}_1)$ studied in
connection with dissipation in the piecewise-linear
model \cite{langer,kl1,crack3,pk}.

These forces define the model.  The equation of motion is simply
\begin{equation}
\ddot u_{\vec{x}} = \sum_{\vec{x} ' \in nn} \left(
f_{\vec{x},\vec{x}'} + g_{\vec{x},\vec{x}'} \right)
\end{equation}
We shall primarily be interested here in steady-state cracks,
where the displacements have the Slepyan form
$u_{x,y}(t)=u_y(t-x/v)$.  Furthermore, we will focus on symmetric
cracks, such that $u_{-y}(t)=-u_y(t+1/(2v))$, which is the symmetry
appropriate to steady-state cracks in the piecewise-linear
model~\cite{mg}. Then, solving for the steady-state crack means
solving for $N$ functions of time, characterizing the time
development of a typical mass point in each row of the lattice.
The equations are non-local in time, due to the coupling between
different lattice points. As there is no hope of constructing
analytic solutions at general $\alpha$, we solve this problem
numerically.

To proceed, we discretize the time variable in units of some small
$dt$. The key insight involved in constructing a numerical
procedure is noting that the steady-state equations have a banded
structure, just as was the case for the piecewise-linear
model~\cite{kl1}. This is due to the fact that mass points are
coupled to nearest neighbors, which gives rise via the Slepyan
ansatz to a coupling to displacements at a finite time separation. Of
course, the equations here are everywhere nonlinear and we need to
employ Newton's method to find a solution. This is nonetheless
computationally feasible, since the update step in Newton's method
is a linear problem with the aforementioned banded structure.

Imagine searching for a solution assumed to have some velocity
$v$. There is one equation of motion for each displacement field
$u_y$ at each discrete instant in time.  In addition, fixing the
translation invariance by some condition such as
$u_{\sqrt{3}/4}(0)=1/2$ gives one additional equation, which we
can use to solve for the driving displacement $\Delta$. The only
problem is that $\Delta$ is involved in many different equations
and so destroys the banded structure of the problem. Two different
ways to circumvent this difficulty present themselves. The simpler
of the two is to guess, for a given velocity, a value of $\Delta$,
and solve a modified system of equations where the equation of
motion at $t=0$, $y=\sqrt{3}/4$ has been dropped, and the time
translation has been fixed.  The resulting banded problem can be
solved economically, and the violation of the dropped equation of
motion calculated. This defines a mismatch function, which has a
zero at the true value of $\Delta$, which is located by a standard
zero-finding routine.  A more efficient approach is to realize
that solving the entire linear system can be accomplished at
essentially no more expense than the fully banded modified system
of the first approach.  The algorithm for achieving this is
presented in Appendix A.  The advantage of this approach is that
the Newton solver produces a value of $\Delta$ directly, without
the need of invoking a outer root-finding routine.

\section{The small velocity regime}

We have already seen in \cite{arrest} that smoothing the form of
the force-law effects a dramatic reduction in the window of
drivings for which arrested cracks exist.  As the moving crack
solution arises as a backward bifurcation of the arrested crack,
we can expect that small velocity cracks are also extremely
sensitive to the smoothness of the force-law. An indication of
this can be found in our previous study \cite{kl1,crack3} of a
continuous-$x$, discrete-$y$ model, which did not have any window
of arrested cracks.  There the velocity rose linearly from zero as
the driving $\Delta$ increased from the Griffith value $\Delta_G$,
with a slope inversely proportional to $\eta$. The study of small
velocity solutions with sharp but not discontinuous force laws
should be directly related to experimental data on the onset of
propagating cracks in materials such as single crystal
silicon~\cite{silicon}.

\begin{figure}[h]
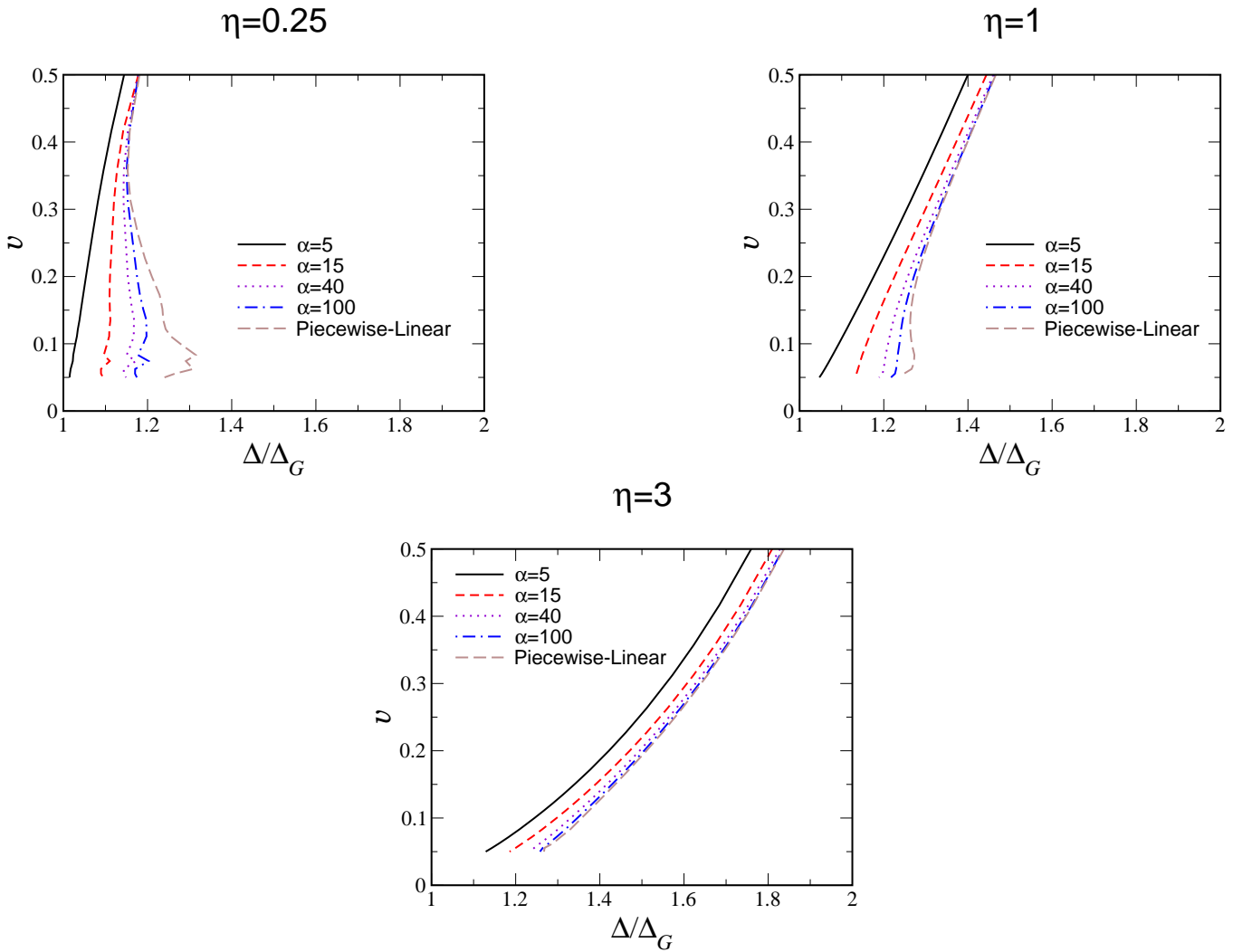

\includegraphics[width=2.8in]{fig1a.eps}\hfill\includegraphics[width=2.8in]{fig1b.eps}
\centerline{\includegraphics[width=2.8in]{fig1c.eps}}
\caption{Dependence of $v$ on $\Delta/\Delta_G$ for $\alpha=5$, $15$, $40$, and $100$ together with the piecewise-linear model for $\eta=0.25$, $1$ and $3$. $N=3$, $dt=0.1$.}
\label{fig1}
\end{figure}

As mentioned above, we solve the steady-state problem via Newton's
method. We present in Fig. \ref{fig1}(a-c) a graph of $v$ versus
$\Delta/\Delta_G$ for various values of $\eta$ and $\alpha$,
together with the results from the piecewise-linear limit.  We
chose to present data for $N=3$ so as to be able to investigate
smaller velocities at reasonable computational cost; the small
velocity regime is relatively insensitive to $N$ \cite{kl1}. A few
general trends are evident from these plots. First, the effect of
$\alpha$ is much more pronounced at small velocity. Second,
smoothing the potential by decreasing $\alpha$ postpones the
onset of the backward branch of the curve to lower velocity.  This
is consistent with the narrowing of the window of arrested cracks
with decreasing $\alpha$, since the curve is forced to turn back
to meet the end of the arrested crack branch.  Decreasing $\alpha$
also decreases the amplitude of the oscillations present at small
$\eta$. Examining the effect of varying $\eta$, we see that the
larger $\eta$'s are less sensitive to $\alpha$.   This is to be
expected, since even in the piecewise-linear model $\eta$ reduces
the extent of the backward branch which is one of the primary
consequences of having smaller $\alpha$. This reduction is related
to the increase in the size of the process zone with increasing
$\eta$~\cite{crack3}, which smoothes out the lattice structure in
a similar manner to that accomplished by smoothing the potential.
Thus, in general, decreasing $\alpha$ and increasing $\eta$ both
act to reduce the lattice-trapping effects which give rise to the
backward branch.  It should be noted that this analysis is consistent
with the onset of running cracks directly above $\Delta_G$ in a 
molecular dynamics simulation using Lenard-Jones potentials\cite{paskin}.

\begin{figure}
\centerline{\includegraphics[width=3.25in]{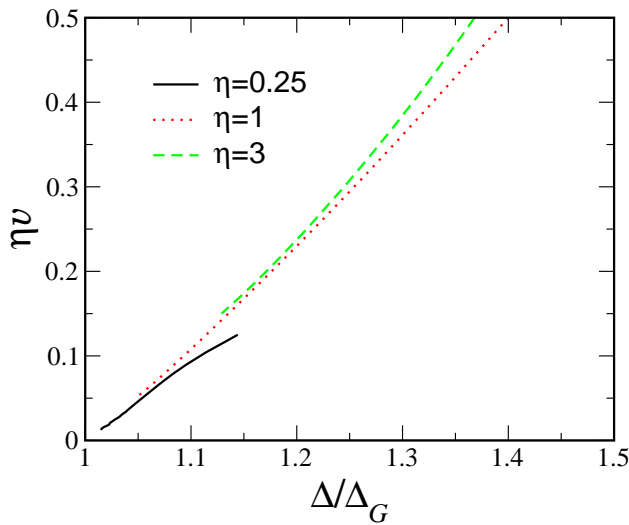}}
\caption{Dependence of $\eta v$ on $\Delta/\Delta_G$ for $\alpha=5$, for
$\eta=0.25$, $1$, and $3$. Again, $N=3$, $dt=0.1$.}
\label{fig2}
\end{figure}

Another way to see that reasonably smooth force-laws give rise to
essentially $x$-continuum behavior is to look at $\eta v$ as a
function of $\Delta$ for different $\eta$'s. In Fig. \ref{fig2}, we present
the data for the case $\alpha=5$.  We see that the data overlap
extremely well, except for the highest velocity data for each
value of $\eta$.  This is an indication that, as in the
$x$-continuum problem~\cite{kl1}, the only velocity scales are
$\eta$ and the wave speed, $c=\sqrt{3/2}$. For the
piecewise-linear model, on the other hand, where the $x$-lattice
scale is important, the $\eta v$
scaling~\cite{sander} only sets in for large $\eta$, where the
process zone is large \cite{crack3}.

\begin{figure}
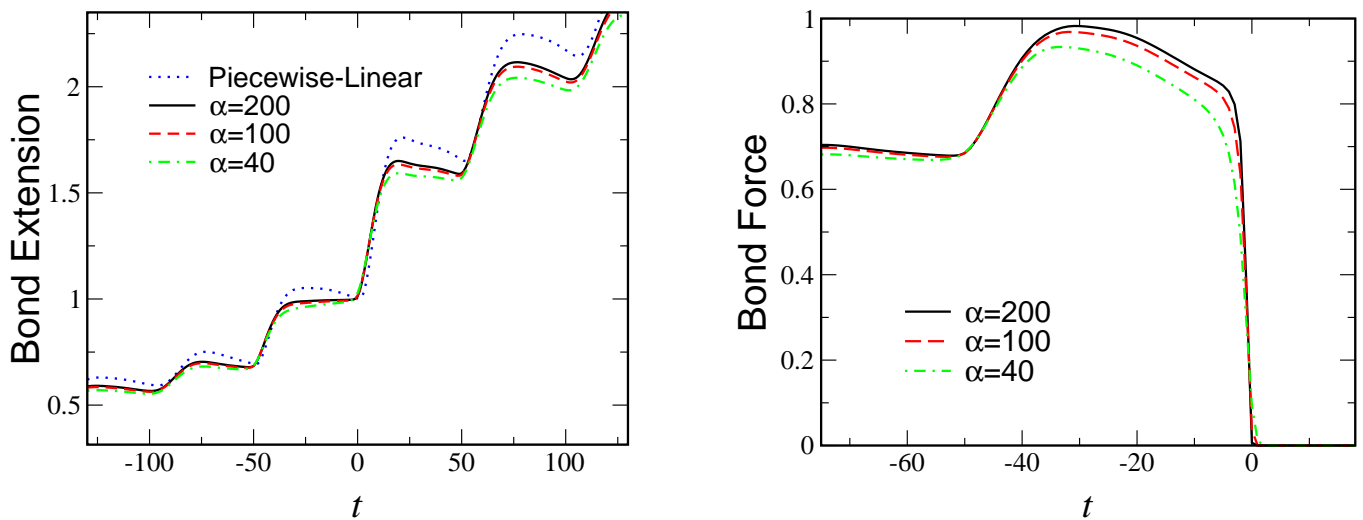

\includegraphics[width=3.25in]{fig3a.eps}\hfill\includegraphics[width=3.25in]{fig3b.eps}
\caption{a) Bond extension along crack surface versus $t$, for
$\alpha=40$, $100$ and $200$ together with the piecewise-linear
model results. b) Bond force along crack surface versus $t$, for
$\alpha=40$, $100$ and $200$. In both figures, $N=3$, $\eta=0.25$,
$v=0.1$ and $dt=.1$.}
\label{slow_bond}
\end{figure}

The other point of interest is the nature of the solutions for
small $\eta$. In the piecewise-linear model, the solutions on the
backward branch are inconsistent at small $\eta$, since the
underdamping of the forward running waves leads to pre-cracking.
This is of course not a problem in our fully nonlinear model,
since cracking here is a reversible process. Nevertheless, it is
amusing to note that the solutions on the backward branch for
small $\eta$ do not behave in this manner.  They do not crack and
reheal.  Rather, they exploit the smooth transition in the force
law to crack in a slow, monotonic fashion.  We show in Fig. \ref{slow_bond}a
the extension of the bond along the crack surface as a function of
time, for $\alpha=40$, $100$, and $200$, along with the analogous
result for the piecewise-linear model.  We see that the limit of
large $\alpha$ does not correspond to the behavior of the
piecewise-linear model. Not surprisingly, then, it turns out the
$v$ vs. $\Delta$ curve for large $\alpha$ does not converge to the
piecewise-linear result.  Thus, for example, for the case
considered in Fig. \ref{slow_bond}a, namely N=3, $v=.1$, $\eta=0.25$, the
limiting value of $\Delta /\Delta_G$ for large $\alpha$ is 1.2140,
as compared to 1.2778 for the piecewise-linear model. Whereas the
bond extension surpasses the critical extension of unity in the
piecewise-linear model, and then returns to unity before cracking,
the bond extension in the nonlinear model is monotonic, and spends
a long time near critical extension.  A clearer picture of this
emerges in Fig. \ref{slow_bond}b, 
where we present the force exerted by the bond
as a function of time for the nonlinear model.  We see that the
bond exploits the nonlinearity of the force-law to crack in a very
slow fashion.

\section{The large velocity regime}

It is known~\cite{marderliu,mg} that the piecewise-linear solution
is inconsistent at large velocity due to the cracking of other
bonds. This has been extensively studied for the Mode III problem
in an infinite square lattice and well as for Mode I in
a triangular lattice\cite{pk}.  We thus expect that for
velocities larger than the critical one at which the inconsistency
first appears, the solutions of our nonlinear model will diverge
significantly from those of the piecewise-linear model.

\begin{figure}[b]
\centerline{\includegraphics[width=3.25in]{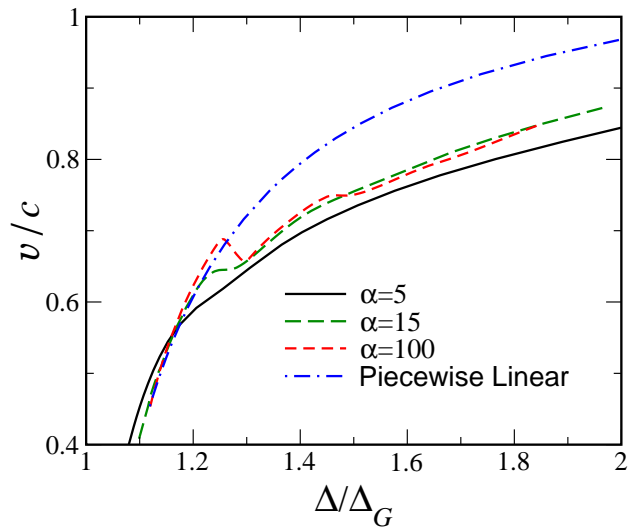}}
\caption{$v/c$ versus $\Delta/\Delta_G$ for $N=8$, $\eta=0.1$. Data
is presented for the cases $\alpha$=5, 15, and 100, and well as for
the piecewise-linear model.}
\label{hi_v}
\end{figure}

To begin, we use our steady-state code to find large velocity
solutions of the equations of motion. We present in Fig. \ref{hi_v} data
for $v$ versus $\Delta/\Delta_G$ at three values of $\alpha$,
together with the analogous results from the piecewise-linear
model. We see that for velocities less than about $0.6 c$, the
curves basically coincide, with the smallest $\alpha$ curve lying
slightly below the others. However, in the range $0.6c < v < 0.7c$,
the data exhibit a marked sensitivity to $\alpha$.  The
low $\alpha$ curve is flat in this region, and as $\alpha$ is
increased, monotonicity is destroyed and a local maximum and
minimum are created.  These features are completely absent in the
piecewise-linear data, indicating that the additional bond
breaking, absent by construction in the piecewise-linear solution,
is responsible. Clearly, once a local maximum in $v$ versus
$\Delta$ sets in, as it does for large enough $\alpha$, the
steady-state solutions are unstable for $\Delta$'s past the
maximum.  We address this issue in the next section, where we
perform a stability analysis of the steady-state solutions.
Nevertheless, it is interesting to note that past this band of velocities,
the dependence on $\alpha$ is again very
mild, but very different from the piecewise-linear results.

\section{Stability Analysis}
Next, we consider the linear stability of the traveling
wave state given by the aforementioned Slepyan ansatz
$u_{x,y} (t) = u_y (t-v/x)$. Specifically, we assume a solution of
the form
\begin{equation}
u_{x,y} (t) \  = \ u_y^{(0)} (\tau ) + e^{-\omega x} \delta u_y (\tau )
\end{equation} where $\tau$ is the traveling wave coordinate $t-x/v$
and $u^{(0)}$ is the previously determined solution.
For the stability problem, we take $\delta u << 1$ and expand to linear order.
Note that the last term can be written in the alternate form
$ e^{\omega v t } \tilde{\delta u_y} (\tau )$
with \begin{equation}
\tilde{\delta u_y} (\tau ) = \delta u_y (\tau ) e^{v \omega \tau}
\end{equation}
Hence, for stability we must have {\em Re} $\omega <0$, i.e. perturbations must decay in time
at a fixed position in the frame moving with the crack tip.

We substitute this assumption into the equation of motion. This leads to
the linear problem
\begin{eqnarray}
\frac{d^2}{d \tau ^2 } \delta u_y (\tau ) & = &
\sum_{nn} G(\vec{x},\vec{x}') \ \left( \delta u_y (\tau ) -
\delta u_{y'} (\tau ')
e^{\omega (x-x')} \right) \nonumber \\ && +
\eta k_{\vec{x},\vec{x}'} \left( u_y^{(0)} (\tau ) -
 u_{y'}^{(0)} (\tau  ') \right)  \frac{d}{d\tau} \left( \delta u_y (\tau ) -
\delta u_{y'} (\tau  ') e^{\omega (x-x')} \right)
\end{eqnarray}
where we have defined
\begin{equation}
G(\vec{x},\vec{x}') = f' \left( u_y^{(0)} (\tau ) - u_{y'}^{(0)} (\tau  ') \right)
+ \eta  k'_{\vec{x},\vec{x}'} \left( u_y^{(0)} (\tau ) - u_{y'}^{(0)}
(\tau  ') \right) \
\frac{d}{d\tau} \left( u_y^{(0)} (\tau ) - u_{y'}^{(0)} (\tau  ') \right)
\end{equation}
and of course $\tau ' = t- x'/v$.

To find the allowed values of $\omega$, we proceed as follows.
First, we impose the asymptotic boundary conditions that at all
$\vec{x}$ outside of our explicit lattice points, $\delta u = 0$.
We then pick an arbitrary normalization by fixing $\delta
u_{y=\sqrt{3}/4} (0) =1$ and simultaneously relax the equation of
motion at that same point. This procedure, implemented for some
guessed value of $\omega$,  converts our problem into an
inhomogeneous, banded system of linear equations for the complex
variables $\delta u_y (\tau )$ which can be solved by standard
techniques. The missing equation then forms a complex mismatch
function whose zeroes then determine the allowed values of the
eigenvalue $\omega$.

Two other details should be noted. First, we do not impose any symmetry
restrictions on our perturbation with respect to reflections about the
crack plane. Also, any time we find a root with some specific value of {\em Im}
$\omega$, an equivalent solution exists with
\begin{eqnarray}
\text{\em Im} \ \omega & \to & 2\pi -\text{\em Im} \ \omega \nonumber \\
\delta u_y & \to & \delta u_y ^* \ \ \  \text{\em (even rows)} \nonumber \\
\delta u_y & \to & - \delta u_y ^*\ \ \ \text{\em (odd rows)}
\end{eqnarray}
By even and odd, we mean the parity of the lattice units of distance from the row
$y=\sqrt{3}/4$.
This result follows from the fact that the governing equation is real
and that the $x$ spacing of nearest neighbor points on the rows
separated by an even number is $\pm 1$, whereas it is $\pm
\frac{1}{2}$ for odd rows.
A corollary of this is that a root at exactly {\em Im} $\omega = \pi$ has an
eigenfunction which is purely real on even rows and purely imaginary on
odd ones.

One obvious test of our stability analysis arises from the fact that
translation invariance guarantees that there is a root at $\omega =0$
with eigenfunction
\begin{equation}
\delta u_y (\tau ) \ = \ \frac{d}{d\tau} u_y ^{(0)} (\tau ) /
\frac{d}{d\tau} u_{y=\sqrt{3}/4} ^{(0)} (0)
\end{equation}
Actually, our discretiztion for numerical purposes of the continuous variable
$\tau$ breaks this exact invariance and hence the mode should be at zero
only in the small $d\tau$ limit. In the data we present below, the
zero mode is found with a typical accuracy of 10$^{-2}$, and the eigenfunction
to the same accuracy agrees with its expected form.

\begin{figure}
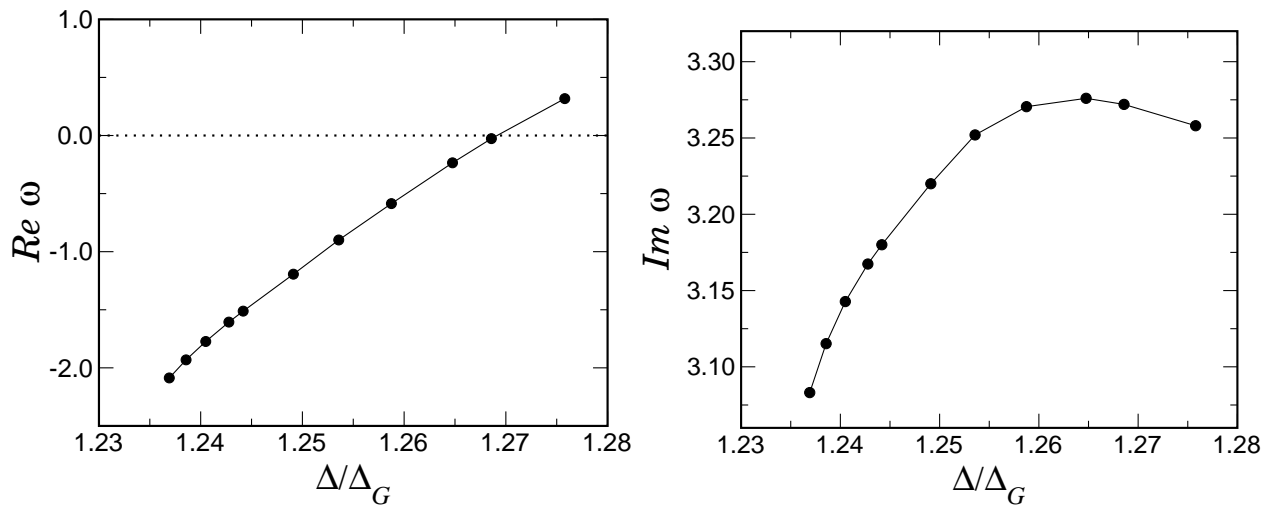

\centerline{\includegraphics[width=3.25in]{fig5a.eps}
\includegraphics[width=3.25in]{fig5b.eps}}
\caption{Eigenvalue for stability problem, with N=8, $\alpha =15$
and $\eta =.1$. The calculation was done with $L=200$ and
$n_b$=12; note that by definition $v=1/(n_b dt)$. {\em Re} $\omega$ is
shown in (a), {\em Im} $\omega$ in (b). } 
\label{fig5}
\end{figure}

In Fig. \ref{fig5}a,
we show the basic result that emerges from our analysis, namely
that as the driving displacement is increased, there is a mode which
crosses the stability threshold. In Fig. 5b, we show the associated
$Im \omega$. At the point of instability,  $Im \omega$ is close
to but not exactly equal to $\pi$. Were it equal to $\pi$, this would be
a period-doubling instability, as the perturbative displacements
at neighboring $x$ values (at the same $\tau$) would alternate in sign.
More general values of Im $\omega$ signify a Hopf bifurcation which
in general has a frequency incommensurate with the original frequency
associated with moving one lattice spacing. The signature of such a
bifurcation should be a slowly oscillating crack tip speed, with the
oscillation frequency $\omega _{tip} = |Im \ \omega - \pi|$.

\begin{figure}
\centerline{\includegraphics[width=4.25in]{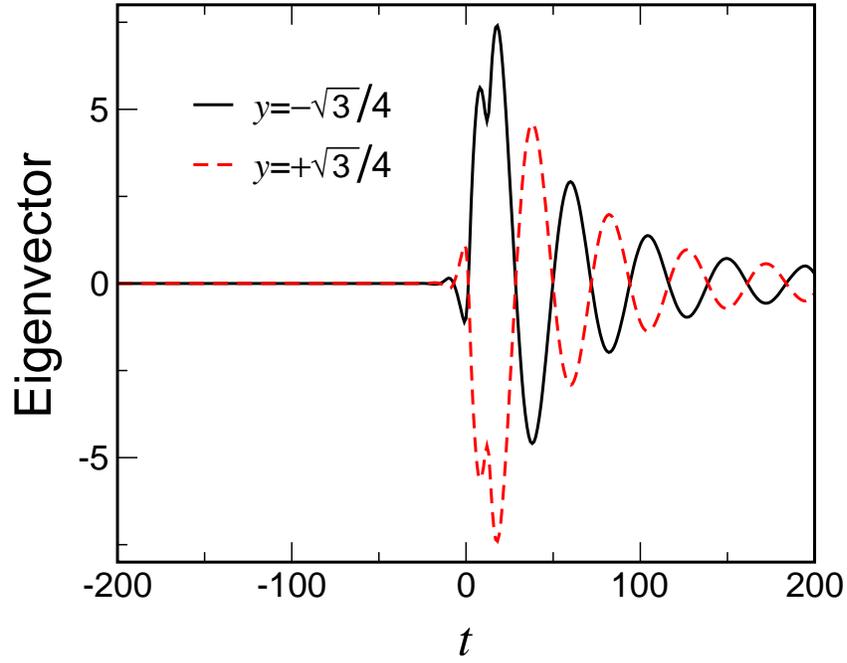}}
\caption{Eigenvector for stability problem, with N=4, $\alpha =50$ and
$\eta =.2$. The calculation was done with $L=200$ and $n_b$=12, and at
the point of instability, dt = .08702.}
\label{fig6}
\end{figure}

In Fig. \ref{fig6}, we show the real part of the eigenfunction for the
rows $y=\pm\sqrt{3}/4$ as a function of the traveling wave
coordinate $\tau$. Note that the perturbation is almost, but not
exactly, antisymmetric around the crack plane. The eigenvector decays
slowly downstream and very rapidly upstream of the crack tip; it
is therefore a localized mode connected with the tip dynamics diverging
from the pure Slepyan form.

\begin{figure}
\centerline{\includegraphics[width=4.25in]{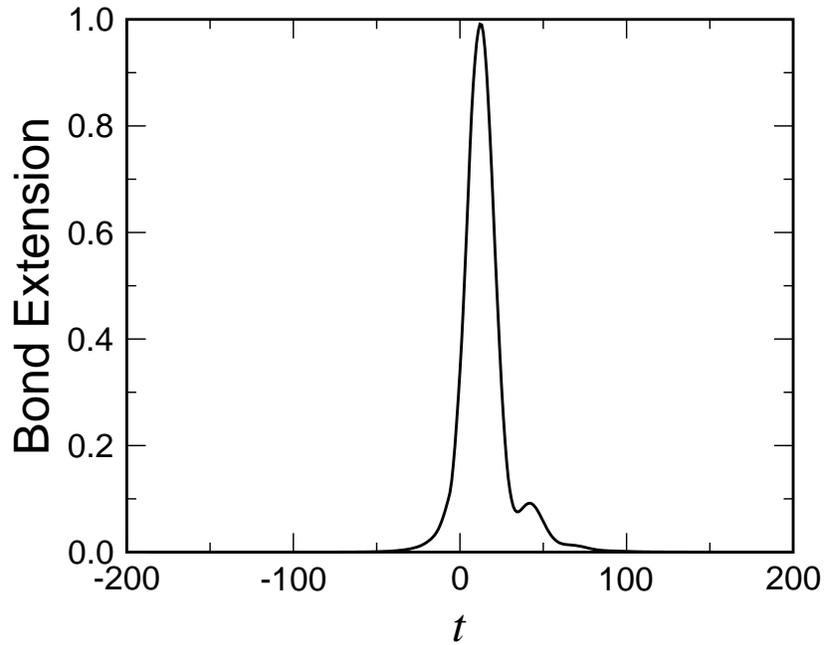}}
\caption{Horizontal bond extension for marginally stable solution,
with N=4, $\alpha =50$ and $\eta =.2$. The calculation was done
with $L=200$,  $nb= 12$, and $dt$ = .08702.}
\label{fig7}
\end{figure}

In previous studies of piecewise-linear models (which correspond
to the limit $\alpha \rightarrow \infty$), it was noted that the
Slepyan solution breaks down above a critical velocity. There, the
criterion for this breakdown was connected to the fact that
certain bond extensions other than those along the presumed crack
line go above the breaking threshold. In our current model where
the force-law is analytic, there is no such criterion and the
steady-state solution found by our procedure does not need to be
checked for any auxiliary condition. On the other hand, we do
observe a linear instability. It is therefore of interest to
compare these two criteria, i.e. to investigate whether the
stability criterion in the large $\alpha$ limit goes over smoothly
to the inconsistency criterion for the ideally brittle case. To do
this, we have plotted in Fig. \ref{fig7} the horizontal bond extension
along the crack line for a steady-state solution at $\alpha=50$
right at the onset of the linear instability. Note that the bond
extension is extremely close (roughly of order 1/$\alpha$) to
what would be the breaking threshold. Thus, the results of the
ideally brittle case as to where nontrivial spatio-temporal
dynamics of the crack tip sets in are not an artifact of the
force-law discontinuity; i.e., smoothing the spring law does not
alter the basic conclusion. This is good news, as the ideal case
has proven rather amenable to analytic techniques which enable
calculations to be done for much larger $N$ (even for infinite
$N$) than is possible for any direct numerical scheme.

\begin{figure}
\centerline{\includegraphics[width=4.25in]{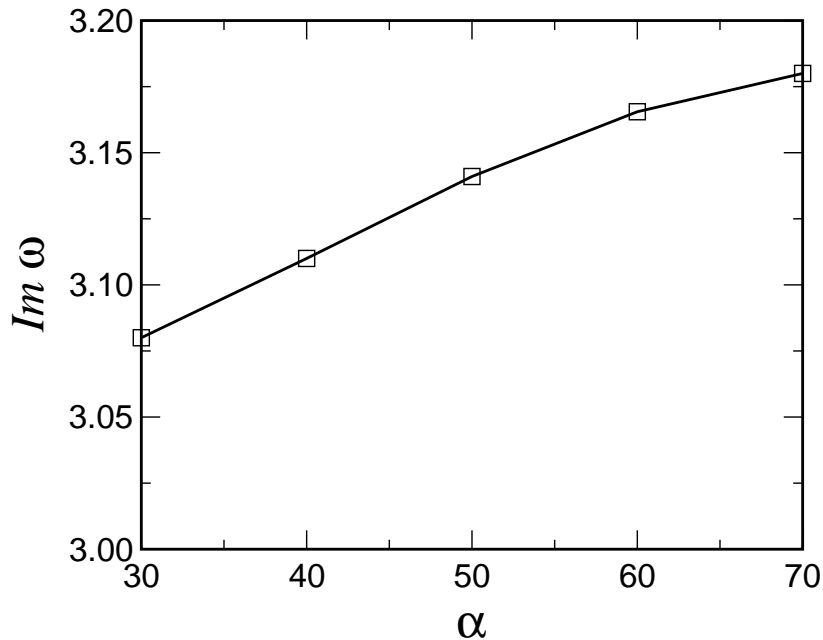}}
\caption{Variation of Im $\omega$ at the marginal stability point with $\alpha$;
data is all for N=4, $\eta=.2$.}
\label{fig8}
\end{figure}

One nagging question posed by our results concerns the fact that
the instability is so close to but not exactly at the
period-doubling point of Im $\omega = \pi$. We have investigated
whether there is any tendency for the mode to get pinned at
the period-doubling point as parameters are varied, as certain
limits are reached etc.. For example, in Fig. \ref{fig8} we show Im
$\omega$ at threshold versus $\alpha$. As far as we have been able
to determine, this pinning does not occur and the generic
bifurcation always leads to an incommensurate oscillating tip
state. We will compare this prediction with direct numerical
simulations of the equations of motion in the next section.

\section{Comparison to Simulation}

In order to test the results of the stability analysis, and also
to investigate the nature of the dynamics past the instability
threshold, we have implemented a direct numerical simulation of
the equations of motion. Our system is a triangular lattice, of
width $2N+2$ whose top and bottom rows are fixed to have
displacement $\pm \Delta$. We start with a uniformly strained
state, except for the first few left hand columns of mass points.
The initial displacement of these points is taken to vary linearly
in $x$ from $\pm \Delta$ at $x=0$ to the that of the uniformly
strained state at $x=10$. The equations of motion are then stepped
forward in time, using a simple Euler scheme. We track the
velocity of the crack by monitoring the bond extensions, and
noting when they exceed unity, which we take as our criterion of
cracking. Note that in this model, cracking is in principle a
reversible process and so this criterion is merely a convenient
way of keeping track of the dynamics and not an intrinsically
important threshold.

As expected, for moderate $\Delta$ the only bonds that ``crack''
are the diagonal bonds that span the midline. The time between
cracking events is constant, and, to best compare with our
steady-state calculations above, we fix our time step $dt$ in each
case so that there are 6 time steps between each cracking event;
this can then be directly compared to our steady-state solutions
computed with time bandwidth $n_b \equiv 1/(v \, dt)$ =12. We find
that the velocities obtained in this way reproduce extremely well
the velocities calculated by our steady-state code.

As we increase $\Delta$ above the critical value for instability
calculated in the previous section, we indeed find that the pattern of
steady-state cracking breaks down.  Some horizontal bonds on the
crack surface begin to crack, and the average velocity of the crack
falls dramatically, presumably due to the draining of energy by these
extra cracking events.  It is important to note that these broken horizontal
bonds are always {\em behind} the crack tip, in accord with the calculations
of the piecewise-linear model.  It is not clear whether there is in fact
a discontinuous drop in the average velocity.  The shape of the drop does
appear to be insensitive to the $dt$ chosen, so that it in any event does
not appear to be a numerical artifact of finite $dt$.

\begin{figure}
\centerline{\includegraphics[width=3.25in]{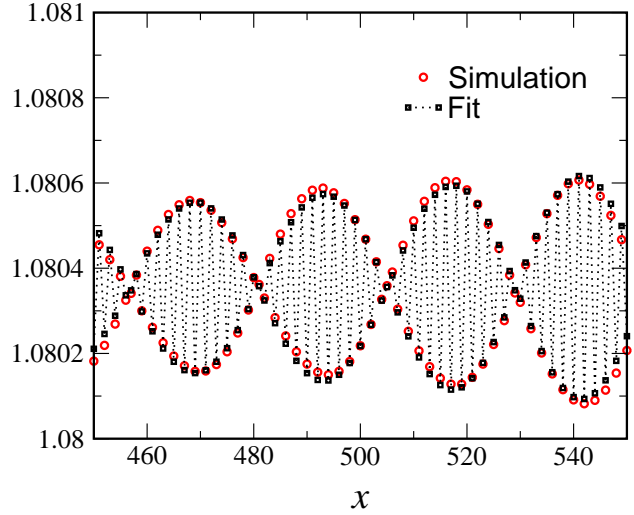}}
\caption{Bond extension at breaking as a function of position $x$ along the
crack surface. Here $N=8$, $\eta=.1$, $\alpha$=15, $\Delta=2.6215$, $\Delta/
\Delta_G=1.2693$,
$dt=0.1053852$.  The fitted curve is $3.57\cdot 10^{-5}\sin(3.27166x -2.85642)
\exp(0.00368x)$.}
\label{hopf_freq-fig}
\end{figure}

As a test of our stability calculation, we have computed the Hopf
frequency directly from the numerical simulation.  We fixed
$\Delta$ to lie slightly higher than the critical value for the
onset of the instability and measured, at the moment of the
breaking of a northeast- southwest diagonal bond, the extension of
the bond.  Due to the discreteness of the time step, this is
somewhat larger than unity. If we had true steady-state
propagation, this value would be constant.  Instead, for this
$\Delta$, we find that the value oscillates about some value, with
the magnitude of the oscillation growing in time. The growth we
associate with the small positive growth rate of the perturbation
in the unstable state, and the frequency of oscillation with the
Hopf frequency. In Fig. \ref{hopf_freq-fig}, we show the data. The
period of oscillation is essentially 1, (in other words, the time
between breakings of this type of bond), which corresponds to a
Hopf frequency of essentially $\pi$.  By numerically fitting a sine
wave to the data, we find that the Hopf frequency is approximately
3.272, in very good agreement with the stability calculation
result shown in Fig. \ref{fig5}b.

\begin{figure}[b]
\centerline{\includegraphics[width=3.25in]{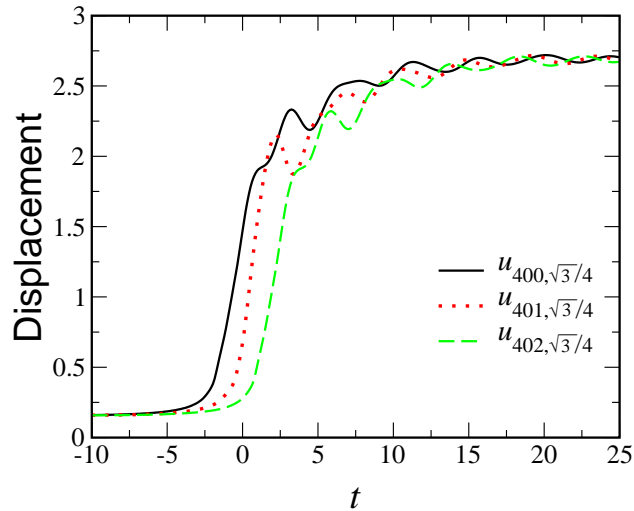}}
\caption{Time trace of displacement of three adjacent points on the upper
crack surface. Here $N=8$, $\eta=.1$, $\alpha$=15, $\Delta=2.7$,
$\Delta/\Delta_G=1.3073$, $dt=0.10858$.}
\label{3trace}
\end{figure}

\begin{figure}
\centerline{\includegraphics[width=3.25in]{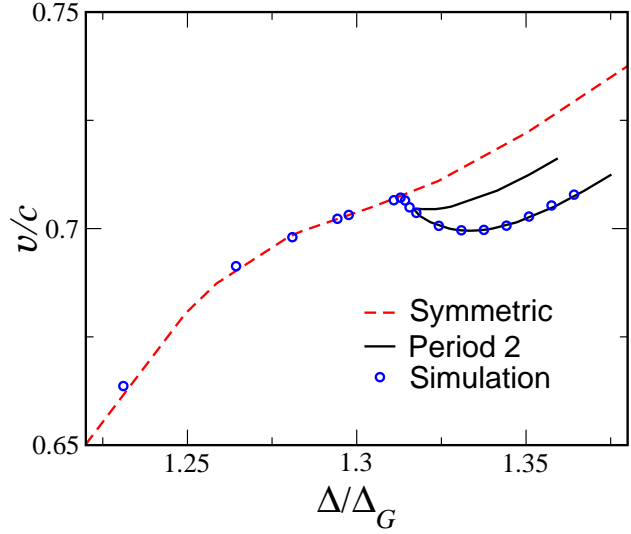}}
\caption{Velocity vs. $\Delta/\Delta_G$ for symmetric and period-2 solutions,
together with measured (time-averaged) velocity from simulation. $N=4$,
$\eta=0.1$, $\alpha=15$.}
\label{bifur}
\end{figure}

As we increase $\Delta$
further, at some point the average velocity starts to increase again.
Examining the solution in this region in more detail,
we show in Fig. \ref{3trace}
the time history of three adjacent points on the crack surface.  We see
that the third trace is identical to the first, and differs significantly
from the second.  It appears then that the system has settled into
a new kind of steady-state solution, more complex than the simple Slepyan
form.  To test this, we implemented a new steady-state algorithm,
designed to allow for this period-2 type solution.  We assumed that
in each row there are two possible time histories for particles, which
alternate.  We also did not assume any symmetry across the crack surface.
Taking care to organize the storage so that the banded structure is maintained,
we employed our banded Newton's method.  Working for the sake of
computational convenience at $N=4$, the algorithm succeeded in obtaining
steady-state solutions where the symmetry between the two time histories
(as well as up-down symmetry across the crack surface) is broken.  We
traced out the velocity-$\Delta$ relation for this solution.  We find
that the line of solutions is disconnected from the symmetric branch
and that in general there are multiple solutions for each $\Delta$.
Starting from a solution obtained from using the simulation to
obtain an initial guess, we tracked the solution as we varied $\Delta$.
The velocity decreased with decreasing $\Delta$ and then turned around
and after some wandering headed off to large $\Delta$ with the velocity
higher than the original branch.  Thus it is the {\em slower} branch
which is physically realized.  This data is presented in Fig. \ref{bifur},
along with the (time-averaged) velocities measured in simulation.

\begin{figure}[b]
\centerline{\includegraphics[width=3.25in]{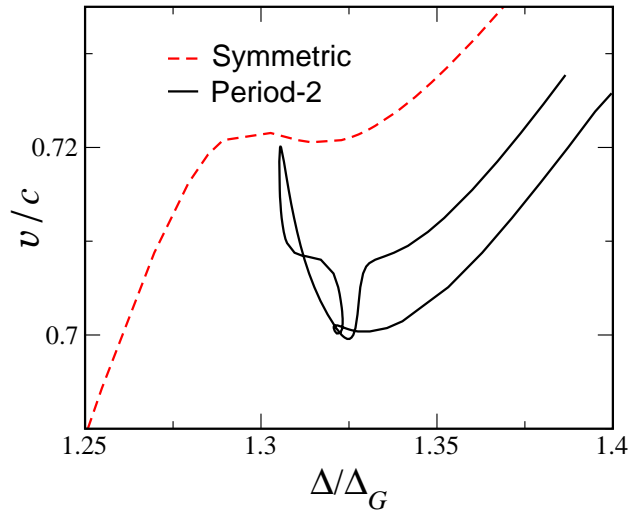}}
\caption{Velocity vs. $\Delta/\Delta_G$ for symmetric and period-2 solutions.
$N=4$, $\eta=0.1$, $\alpha=40$.}
\label{al_40}
\end{figure}

It is reasonable to assume that the existence of a period-doubled
solution very close to the original symmetric branch is associated
with the fact that the Hopf bifurcation is nearly period-doubling.
To check this notion, we note that it was found in the previous
section that the Hopf-frequency passed through $\pi$ for some
particular value of $\alpha$.  For this value, one expects the
period-doubled branch to hit the main branch, corresponding to a
higher co-dimension bifurcation. To address this question, we have
also traced out the bifurcated solution for $\alpha=40$, which is
very slightly below this crossing point. The amazingly baroque
solution curve is presented in Fig. \ref{al_40}, 
together with the
symmetric branch. We see that the bifurcated solution curve turns
up to approach the symmetric curve, before veering away.
Examination of the solutions indicate that the asymmetry is
extremely small in this region of close approach. Presumably,
then, at the crossing value of $\alpha$ the bifurcated solution
actually meets the symmetric branch, and the bifurcation is
perfect at this point. As the meeting is not on the presumably
stable branch of the bifurcated solution, the bifurcation should
prove to be subcritical. A detailed study of the nature of the
bifurcation should prove mathematically interesting, though not
necessarily physically relevant.  It should be noted that we have
also uncovered another, apparently disconnected, branch of
period-2 solutions.  These other solutions presumably play an
important role in the exact nature of the bifurcation, though
again they do not seem to be important for the physics.

\begin{figure}
\centerline{\includegraphics[width=4.25in]{lat_ny=42_delta=5.5_eta=.1.eps}}
\centerline{\includegraphics[width=4.25in]{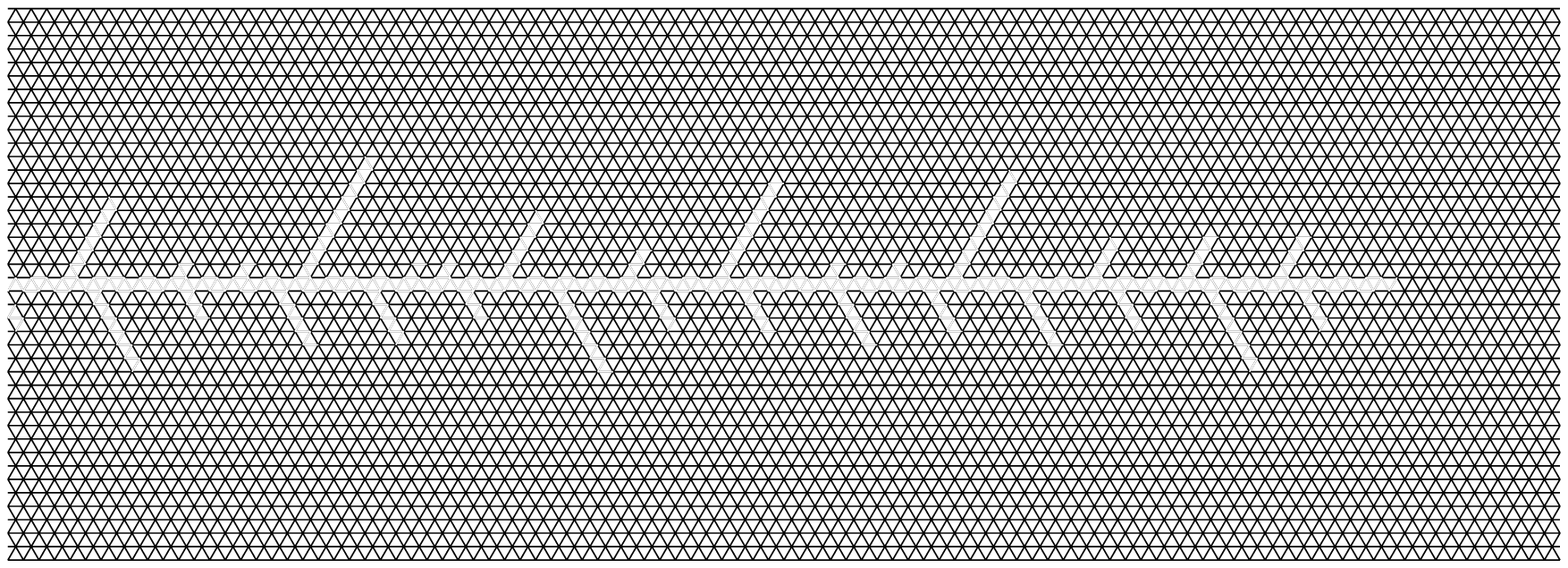}}
\caption{Broken bonds for $N=20$, $\eta=.1$, $\alpha=15$. (Top) $\Delta$=5.5,
$\Delta/\Delta_G$=1.715; (Bottom) $\Delta$=6.5,
$\Delta/\Delta_G$=2.207}
\label{lats}
\end{figure}

Increasing $\Delta$ further, the simulation tracks the bifurcated
steady-state solution until at some point yet additional bonds are
broken, and the velocity falls below that of the bifurcated
solution.  The time-dependence of this state becomes more
complicated than the simple period-2 structure of the bifurcated
steady-state.  It is important to point out that the additional
bonds are also always located behind the crack tip, which remains
on the midline.  Thus the additional bond breaking is reminiscent
of side-branching in dendritic growth~\cite{kkl}; i.e., a
phenomenon induced by the tip and which is left behind by the
growing crack. As opposed to dendritic growth, in Mode III
fracture the sidebranching is generated intrinsically by the tip,
and is not noise-induced~\cite{brener}.  We see, however, no
tendency for the tip itself to leave the midline.  This can be
seen in Fig. \ref{lats}, where the broken bonds are portrayed for
two different values of $\Delta$. Also, the picture shows that as
the driving is increased the size of the sidebranches increases,
but they are always microscopic, growing to a length of about 8 at
the larger $\Delta$. In this latter picture, the crack is moving
at an average velocity of $v/c=.789$, which is quite fast.
Furthermore, the sidebranches are not only short, but the
sidebranching period is also on the lattice scale.  Thus, the
claim of \cite{marderliu} that these sidebranches are related to
the the experimentally seen micro-branching appears to us
doubtful, especially in light of our preliminary work on Mode I
cracking \cite{mode1new}, where the tip dynamics is very
different. This issue is worthy of further exploration.

\begin{acknowledgments}
The work of DAK is supported in part by the Israel Science
Foundation. DAK thanks the Lawrence Berkeley National Laboratory,
where this work was initiated, for its hospitality. HL is
supported in part by the US NSF under grant DMR98-5735.
\end{acknowledgments}

\section{Appendix - Solving the almost banded problem}

In this appendix, we outline how to efficiently solve the linear
problem associated with the Newton's method treatment of the
nonlinear steady-state problem.  It is important to set up the
problem correctly to achieve an almost-banded structure.  We order
the variables with the $y$ index running more quickly, since we
want $N \ll  2L+1$, the length of the system in the $x$ direction.
We order the equations similarly, replacing the equation of motion
for the site $x=0$, $y=\sqrt{3}/4$ (where $x$ runs from $-L$ to
$L$) by the constraint equation $u=u0$, where the constant $u0$
can be chosen arbitrarily. The equation of motion for the site
$(0,\sqrt{3}/4)$ is taken to be the last equation. We also note
that equations of motion have to be taken for the columns $-L-1
\le x \le L-1$, since the number of linear modes which diverge as
$x \to -\infty$ is one greater than the number which diverge for
large positive $x$ (see \cite{kl1,crack3}) due to the presence of
the third derivative operator.

The problem then has the structure
\begin{equation}
\left(\begin{array}{cc} {\cal M}_b & V \\ W^T & 0 \end{array}\right)
\left(\begin{array}{c} \delta U \\ \delta\Delta \end{array}\right) =
\left(\begin{array}{c} A \\ b \end{array}\right)
\end{equation}
Here, ${\cal M}_b$ is a banded square matrix of size $N(2L+1)$
with lower bandwidth $N(n_b + 2)$ and upper bandwidth $N(n_b + 1)$
(with $n_b$ an integer such that the velocity is $v=1/(n_b dt)$).
The replacement of the equation for the distinguished site
$(0,\sqrt{3}/4)$ is necessary to ensure that $\cal M$ is not
singular. The length $N(2L+1)$ vector $V$ contains the derivatives
of the equations of motion with respect to $\Delta$ and the vector
$W$ contains the equation of motion for the distinguished site
(which does not depend on $\Delta$). The shift in the
displacements $\delta U$, and the souurce $A$ are also vectors of
length $N(2L+1)$, and $\delta \Delta$ and $b$ are numbers. We can
explicitly solve this system in terms of ${\cal M}_b^{-1}$, which
is easy to compute due to its banded structure. We find
\begin{eqnarray}
\delta \Delta &=& \frac{W^T{\cal M}_b^{-1} A - b}{W^T{\cal M}_b^{-1} V} \nonumber \\
\delta U &=& {\cal M}_b^{-1} (A - (\delta\Delta) V)
\end{eqnarray}
Hence, our entire system can be solved with no more effort than
would be required for a fully banded problem.


\begin{thebibliography}{99}

\bibitem{review} For a review, see J. Fineberg and M. Marder, Phys. Repts. {\bf
313}, 2 (1999).
\bibitem{fineberg}J. Fineberg, S. P. Gross, M. Marder and H. L. Swinney,
\prl {\bf 67}, 457 (1992); \prb {\bf 45}, 5146 (1992); E. Sharon,
S. P. Gross and J. Fineberg, \prl {\bf 74}, 5096 (1995); \prl {\bf
76}, 2117 (1996).
\bibitem{slepyan}L. I. Slepyan, Doklady Akademii Nauk SSSR {\bf 258},
561 (1981) $[$Sov. Phys. Dokl. {\bf 26},  538 (1981)$]$.
\bibitem{mdsim}S. J. Zhou, D. M. Beazley, P.S. Lomdahl, and B. L. Holian, \prl
{\bf 78}, 479 (1997) ; P. Gumbsch, S. J. Zhou, and B. L. Holian,
\prb {\bf 55}, 3445 (1997); D. Holland and M. Marder, \prl {\bf
80}, 746 (1997).
\bibitem{mg}M. Marder and S. Gross, J. Mech. Phys. Solids {\bf 43},
1 (1995).
\bibitem{kl1}D. A. Kessler and H. Levine, \pre{\bf 59}, 5154 (1998).
\bibitem{crack3}D. A. Kessler, \pre{\bf 61}, 2348 (2000).
\bibitem{pk}L. Pechenik, H. Levine and D. A. Kessler, ``Exact solution for
steady-state mode III cracks in a viscoelastic lattice model",
preprint (1999);
``Steady-state mode I cracks
in a viscoelastic  triangular lattice", preprint (1999).
\bibitem{arrest}D. A. Kessler and H. Levine, \pre{\bf 60}, 7569 (1999).
\bibitem{slepyan2}Sh. A. Kulamekhtova, V. A. Saraikin and L. I. Slepyan,
Mech. Solids {\bf 19}, 102 (1984).
\bibitem{marderliu}M. Marder and X. Liu, \prl {\bf 71},
2417 (1993).
\bibitem{langer}J. S. Langer, \pra {\bf46}, 3123 (1992).
\bibitem{silicon}J. A. Hauch, D. Holland, M. P. Marder, and  H. L.
Swinney, \prl{\bf 82}, 3823 (1999).
\bibitem{paskin}A. Paskin, D. K. Som, and G. J. Dienes, Acta Metallurgica
{\bf 31}, 1841 (1983).
\bibitem{sander} O. Pla, F. Guinea, E. Louis, S. V. Ghasias and L. M. Sander,
\prb {\bf 57}, R13981 (1998).
\bibitem{kkl}D. A. Kessler, J. Koplik and  H. Levine, {\em Adv. Phys.} {\bf 37}, 255-339
(1988).
\bibitem{brener} E. Brener and D. Temkin, \pre{\bf 51}, 351
(1995).
\bibitem{mode1new} D. A. Kessler and H. Levine, in preparation.






\end{thebibliography}
\end{document}